# Evaluating Feasibility of Using Wireless Sensor Networks in a Coffee Crop Through Simulation of Aodv, Aomdv, Dsdv And Their Variants With 802.15.4 Mac Protocol

Ederval Pablo Ferreira da Cruz[1], Luis Eduardo Gottardo[1], Franciele Pereira Rossini[1], Vinicius de Souza Oliveira[1] and Lucas Cellim Pereira[1]

[1]Federal Institute of Espirito Santo (IFES) – Campus Itapina, Colatina-ES, Brazil
`ederval.cruz@ifes.edu.br, luiseduardogotardo@gmail.com,`
`francielypr@gmail.com, souzaoliveiravini@gmail.com,`
`lucascellim@hotmail.com`

## ABSTRACT

*A Wireless Sensor Networks is a network formed with sensors that have characteristics to sensor an area to extract a specific metric, depending of the application. We would like to analyse the feasibility to use sensors in a coffee crop. In this work we are evaluating routing protocols using real dimensions and characteristics of a coffee crop. We evaluate, through simulation, AODV, DSDV and AOMDV and two variants known in this work as AODVMOD and AOMDVMOD with 802.15.4 MAC Protocol. For this comparison, we defined three performance metrics: Packet Delivery Ratio (PDR), End-to-End Delay and Average Energy Consumption. Simulation results show that AOMDVMOD overall, outperforms others routing protocols evaluated, showing that is possible to use WSN in a real coffee crop environment.*

## 1. INTRODUCTION

In the last years several technologies emerged with the objective to assist human being. One of these emerging technologies is known as Wireless Sensor Networks (WSN). The use of new "smart" wireless equipments to sensing and to communicate with each other open new perspectives. In this kind of network, sensors can sensing, measure and gather information from the environment, consequently sending such data to the user.

Wireless sensor networks is an area that several research groups around the world have concentrated their efforts to solve problems in all communication layers, including physical-layer communication up to the development of new applications. Wireless sensor networks consist of many smart sensor nodes, where these sensors are equipped with one or more sensors, a processor, memory, a power supply and a radio wireless channel to communicate with each other [1].

Wireless sensor networks can be considered as a special type of Ad Hoc Networks. They can be static, where the nodes do not have any movement, or mobile, known also in the literature as Mobile Ad Hoc Networks (MANET). Ad Hoc Networks is different from infrastructure networks, where such architecture has a base station coordinating the communication of the nodes. Usually, wireless sensor networks has not centralized control and predefined communication link, transferring signals to the exterior world. All nodes are capable to act as source or sink node at the same time. One of the drawbacks of the wireless sensor networks is due to the fact of the nodes have limited processing power because of their tiny physical size, which limits the capacity of processor and size of battery. When collectively works together, they have an ability to collect information of the physical environment. In the Figure 1 is showed an example of a MANET with infrastructure and infrastructureless or pure ad-hoc manner [2].

One of the main challenges to be solved in Wireless sensor networks is routing data of the source node up to destiny node. Several factors can influence the design of the routing protocols such as: **Node Deployment, Power Consumption, Data Delivery Models, Node/link Heterogeneity, Fault Tolerance, Scalability, Network Dynamics, Transmission Media, Connectivity, Operating Environment, Data Aggregation/Fusion, Quality of Service (QoS), Production Costs, Data Latency and Overhead and Autonomy**. Hence, is possible to see that routing protocols are needed to cope with the nature of wireless sensor networks and that proper routing in ad-hoc networks is the challenge to the designers [3, 4].

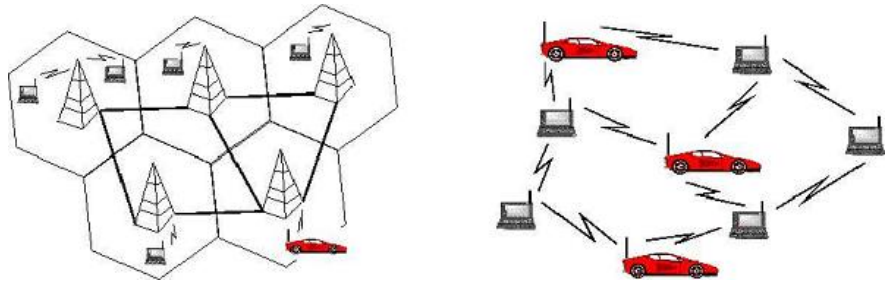

Figure 1. MANET with infrastructure (left side) and infrastructureless (right side) [2]

This work objective to evaluate the feasibility of using Wireless Sensor Networks in a coffee crop through simulations using three classics routing protocols for ad hoc networks and two little variants:

- Proactive routing protocol named DSDV (Destination-Sequenced Distance-Vector Routing Algorithm) [5];
- Reactive routing protocol knows as AODV (Ad Hoc On-Demand Distance Vector Routing) [6]
- Routing protocol based in multiple paths called as AOMDV (Ad hoc On-demand Multipath Distance Vector Routing) [7];
- A modification of AOMDV protocol called in this work as AOMDVMOD. The modification consists to increase HELLO messages dissemination interval to 5 seconds. Default value used by the protocol is 1 second;
- AODV named AODVMOD. The modification consists to increase HELLO messages dissemination interval to 5 seconds. Default value used by the protocol is 1 second.

Furthermore, these routing protocols are simulated using IEEE 802.15.4 as the underlying MAC layer. So, the main contributions of this paper is: (1) evaluate routing protocols for ad hoc networks using realistic dimensions used in a coffee crop and (2) sensors located in the scenarios evaluated considering characteristics used in a real coffee crop such as spacing between coffee seeds.

The remaining of the paper is organized as follows: In the Section 2 is described about the routing protocols used in this work. Section 3 shows the related works. Section 4 describes about the performance evaluation describing scenarios, communications patterns, metrics and more. In the Section 5 the results are analysed and discussed and the paper in concluded in the Section 6.

## 2. ROUTING PROTOCOLS BACKGROUND

The ad hoc networks routing protocols have as main objective to find better route between a source-destiny node by the information existing in the routing tables of the nodes. In the WSN, the movement of the nodes is not a challenging considering the type of application to deploy, and, commonly, when movement is used, the nodes have lower speed than other kind of ad hoc networks (e.g. VANETs). One of the major restrictions to be considered in the WSN is the battery of the nodes. Sensor nodes have limited available power. So, energy efficient routing protocols are truly crucial for life of WSN. But energy consumption is not the only factor to be analysed. Considering that in a real environment, such as coffee crop (our objective to deployment WSN), user needs information more fast possible (low delay) and with reliability (high packet delivery rate). So, is possible to see that other metrics are very important as well.

Routing protocols in WSN can be classified depending of the characteristics such as [3]:

- Route Establishment
- Network Structure
- Protocol Operation
- Initiator of Communicator

The Figure 2 gives and overview of the routing taxonomy.

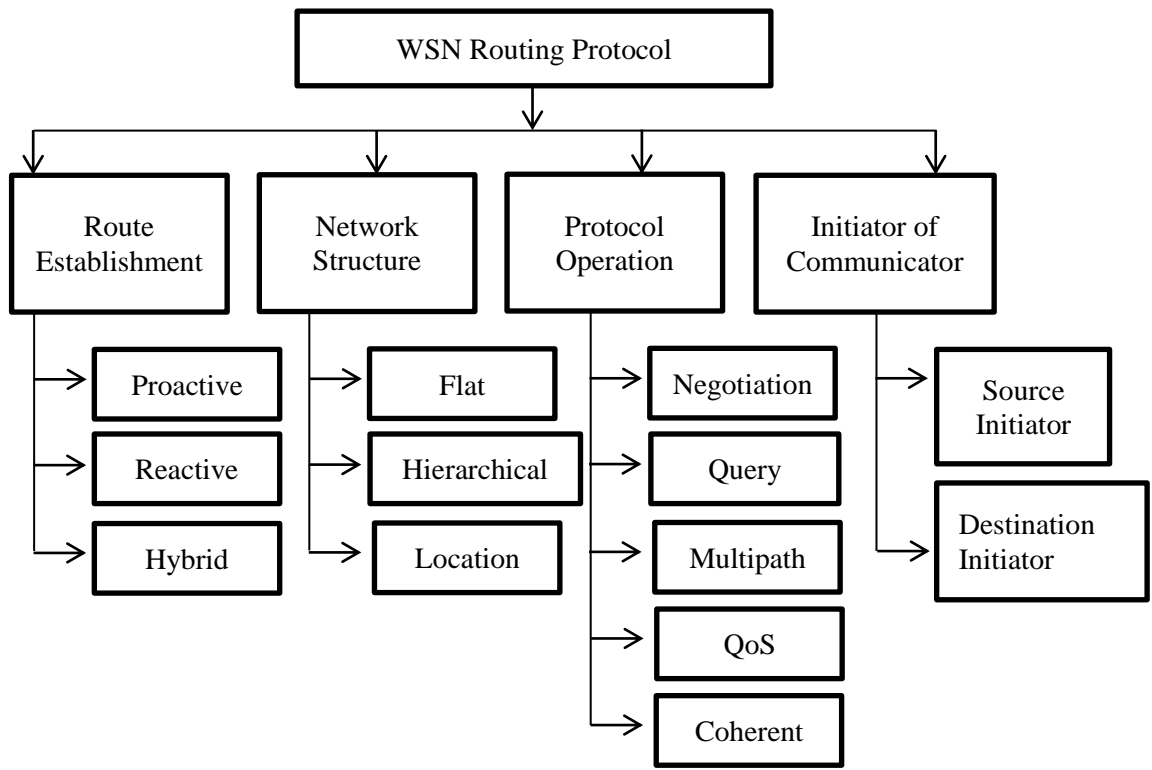

Figure 2. Taxonomy of Wireless Routing Protocol

Considering our work, we will focus in the Route Establishment classification. More details about the protocols evaluated inserted in such classification will be describe later.

## 2.1 ROUTE ESTABLISHMENT BASED ROUTING PROTOCOL CLASSIFICATION

Different strategies can be used to route packets among source-destiny node in the network. Routing protocols must be developed aiming maximize lifetime network with low delay and high packet delivery ratio. Basically, three different route establishment strategies can be used:

- **Proactive or Table-Driven Routing**: In this approach when the routing protocol begins to work, nodes try to populate its routing tables with possible destination nodes. This kind of network is not suitable for larger networks, as they need to maintain node entries for each and every node in the routing table. This leads to more overhead in the routing table leading to consumption of more bandwidth Classic examples of routing protocols categorized as proactive or table-driven routing are: OLSR [8] and DSDV.

- **Reactive or On-Demand Routing**: This type of routing protocols construct routes only when is necessary to send information. In a situation where a node wants to communicate to another node this kind of protocol searches for the route in an on-demand manner and establishes the connection in order to transmit and receive the packet. A technique usually used by reactive routing protocols to route discovery is using flooding route request packets throughout the network. We can cite as routing protocols in this category: AODV, DSR [9].

- **Hybrid Routing**: In this case routing protocols use both reactive and proactive mechanisms to fill routing tables (create or update new routes). Examples of hybrid routing protocols are**:** ZRP [10]

### 2.1.1 DESTINATION SEQUENCE DISTANCE VECTOR (DSDV)

Each node in the network maintains a routing table for the transmission of the packets. Each node has a list of possible destinations in the network and the number of hops necessary to reach each destination in the routing table. With the objective to maintain the information in the routing tables always updated each station transmits a HELLO message in a broadcast manner periodically. Each new route will contain the following information as showed in Table 1.

Table 1: Fields of Routing Table using DSDV [11]

| DESTINATION | NEXT HOP | HOPS/METRIC | SEQ. NO. | INSTALL TIME | STABLE DATA |
|---|---|---|---|---|---|
|  |  |  |  |  |  |

The newest route is used (identified with the highest sequence number). With this information is possible to identify the old routes from the new ones, thereby avoiding the formation of loops.

### 2.1.2 AD-HOC ON-DEMAND DISTANCE VECTOR (AODV)

AODV (Ad-hoc On-demand Distance Vector) is a loop-free routing protocol for ad-hoc networks based on vector distance algorithm, as DSDV. It is designed to be self-starting in an environment of mobile nodes, withstanding a variety of network behaviours such as node mobility, link failures and packet losses.

The AODV protocol is based in topology information that works in a reactive mode. In other words, when a source node wants to send data to a destination node, a route discovery process is started. In this process, the source node broadcasts a ROUTE REQUEST (RREQ) packet to all your neighbours. Neighbours nodes which do not know an active route to destination node, forward the packet to their neighbors until an available route is found or the maximum number of hops is reached. When an intermediate node knows an active route to the requested destination node, it sends back a ROUTE REPLY (RREP) packet to the source node in unicast mode, enabling to the node creates the route among the source and destination nodes. In case of a link failure, the neighbourhood nodes are notified by route error messages (RERR) on both sides of link.

## 2.1.3 AD-HOC ON-DEMAND MULTIPATH DISTANCE VECTOR (AOMDV)

The main characteristic in AOMDV is computing multiple loop-free paths per route discovery. Due to fact to create multiple redundant paths, the protocol switches routes to a different path when an earlier path fails. Thus a new route discovery is avoided. Route discovery process is initiated only when all paths to a node destination fail. For efficiency, only link disjoint paths are computed so that the paths fail independently of each other. Note that link disjoint paths are sufficient for our purpose, as we use multipath routing for reducing routing overheads rather than for load balancing. The multi-path routing protocol discovers multiple paths during the single route discovery process. These multiple paths can be used for load spreading or as backup routes when the primary route fails [12].

Different of AODV routing protocol, AOMDV do not discard duplicate RREQs. AOMDV look for an opportunity of getting an alternate route with each duplicate RREQ. In AOMDV, RREQ propagation from the source towards the destination establishes multiple reverse paths both at intermediate nodes as well as the destination. Multiple RREPs traverse these reverse paths back, to form multiple forward paths to the destination at the source and intermediate nodes. AOMDV also provides intermediate nodes with alternate paths as they are found to be useful in reducing route discovery frequency. The core of the AOMDV protocol lies in ensuring that multiple paths discovered are loop free and disjoint; and in efficiently finding such paths using a flood-based route discovery. AOMDV route update rules, applied locally at each node, play a key role in maintaining loop-freedom and disjoint-ness properties [13].

## 3. RELATED WORKS

This section presents a brief summary of related works. Chicka et al [14] evaluates AODV and DSR routing protocols for IEEE 802.15.4/ZigBee. The authors aim to analyse, specially, the energy consumption. But the scenario studied in the work does not consider any specific application.

In [15] is realized an evaluation of the network performance of a WSN for Temperature Monitoring in Vineyards. The authors deployed real equipments (called motes), in real city. But in the work the authors do not describe what routing protocol was used in the experiments.

In [16] Santos et al., give an brief overview about feasibility of using WSN in agricultural monitoring where they show that is possible to use WSN with low cost and increasing productivity.

In the work [17] Verona realizes a simulation of a WSN with the objective to use WSN in vineyards with hierarchical or flat approach.

Oliveira et al [18] present in their paper a Quality of Service (QoS) analysis of four Routing Protocols in WSN based on the IEEE 802.15.4 standard in the Monitoring of Wind Farms. They claim that the rising global demand of energy and the scarcity of fossil fuels can be factors that justify to use renewable energies sources in the world, such as wind power.

# 4. PERFORMANCE EVALUATION

This section presents the evaluation of the AODV, AODVMOD, AOMDV, AOMDVMOD and DSDV routing protocols under 802.15.4 MAC protocol in a scenario that aiming to represent a coffee crop scenario. We adopted 802.15.4 because this MAC protocol has some characteristics such as low transmission rate and consumed energy when compared with 802.11, and the devices that use 802.15.4 as MAC protocol are cheaper. This shows that in our case, aiming to use sensors in a coffee crop, 802.15.4 is more viable to adopt than 802.11. In the Figure 3 is showed (in the selected area) the dimensions of the real area deployed in the simulator. In the Table 2 is showed the coordinated points extracted from Google Earth of the scenarios used in our simulations.

Table 2: Coordinated points of the area used in the simulations

|   | Latitude | Longitude |
|---|---|---|
| A | 19°29'48.16"S | 40°45'32.54"W |
| B | 19°29'46.86"S | 40°45'31.95"W |
| C | 19°29'48.34"S | 40°45'25.16"W |
| D | 19°29'49.62"S | 40°45'25.75"W |

To realize the simulations, NS-2 simulator [19] was adopted to evaluate behaviour of the wireless sensor networks deployed in this work. NS-2 is a discrete event simulator developed by the VINT project research group at the University of California at Berkeley. There are some interesting characteristics such as: (a) node mobility, (b) a realistic physical layer with a radio propagation model and more.

We consider 3 different scenarios, where the number of sensors varies as follows:

- Scenario 1: 40 nodes + 1 sink node
- Scenario 2: 55 nodes + 1 sink node
- Scenario 3: 70 nodes + 1 sink node

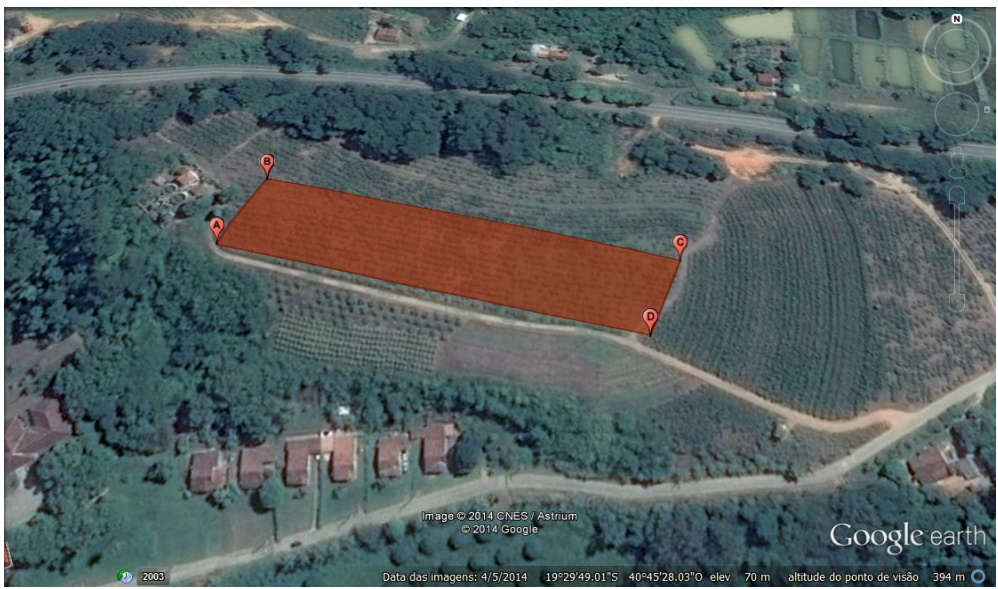

Figure 3. Selected area deployed in the NS2 simulator extracted from Google Earth.

It is important to mention that the sink node (represented in the Figures 4, 5 and 6 by the white circle) is located in the center of the scenario evaluated. To give more details about the arrangement of the nodes in the scenarios evaluated, are presented scenarios 1, 2 and 3 in the Figures 4, 5 and 6, respectively.

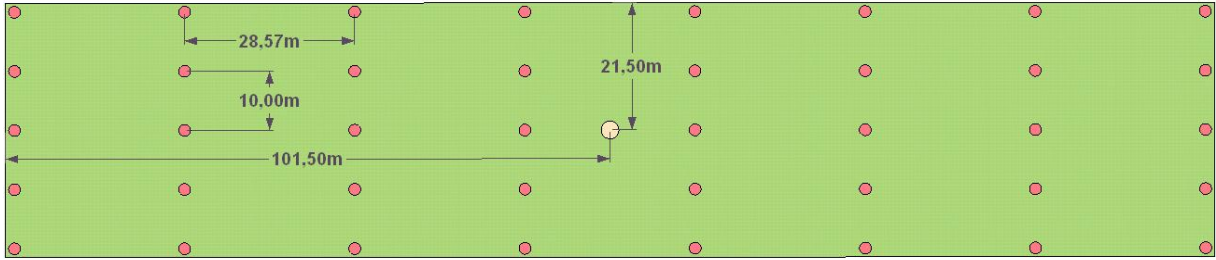

Figure 4. Scenario 1 deployed in the NS2 simulator.

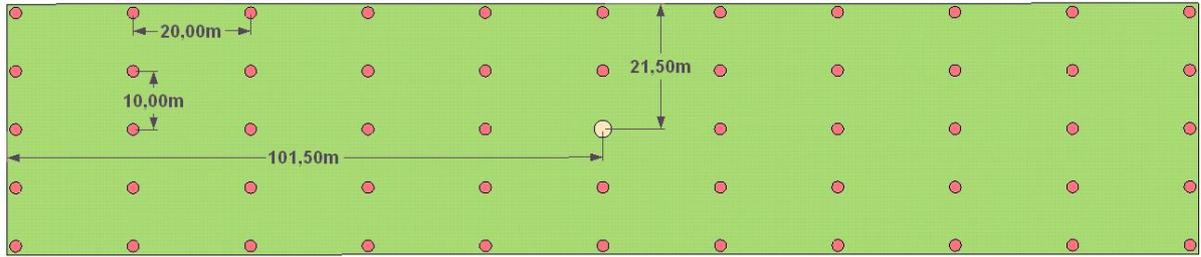

Figure 5. Scenario 2 deployed in the NS2 simulator.

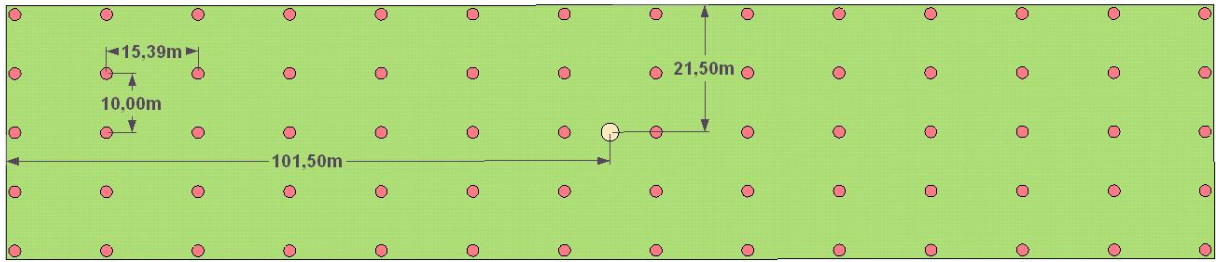

Figure 6. Scenario 3 deployed in the NS2 simulator.

Table 3 summarizes the parameters used in the employed simulation.

Table 3: Parameters used in the simulation

| MAC Protocol | 802.15.4 |
| --- | --- |
| Radio Propagation Model | Two Ray Ground |
| Routing Protocols | AODV, AOMDV, AODVMOD, AOMDVMOD, DSDV |
| Simulation Time | 180 seconds |
| Number of Nodes | Scenario 1: 40 + 1sink |
| | Scenario 2: 55 + 1 sink |
| | Scenario 3: 70 + 1 sink |
| Packet Size | 512 bytes |
| Transmission Rate | 4 packets/sec |
| Traffic Type | CBR |
| Initial Energy | Sink Node – 50 kJ |
| | Other nodes – 5 kJ |
| Transmission Range | 60 meters |

On Table 4 is showed the relation between the amount of source nodes and connections created for each scenario. We adopted such characteristic considering that in a real communication is possible that one source node can create several connections simultaneously (2 or more different nodes). The number of connections in each scenario is equal to 20% of the total of nodes (except sink node) of the scenario.

Table 4: Connections and source patterns used in the simulations for each scenario

| **Amount of connections for each scenario** | |
|---|---|
| 40 nodes + 1 sink | 6 (different sources) and 8 (connections) |
| 55 nodes + 1 sink | 7 (different sources) and 11 (connections) |
| 70 nodes + 1 sink | 10 (different sources) and 14 (connections) |

It is important to mention that the results presented in this paper are averaged of 30 simulations for each scenario and for each protocol. All simulations results presented were obtained guaranteeing a 95% confidence interval.

### 4.1 PERFORMANCE METRICS

As already showed in the Tables 3 and 4, the performance of the routing protocols AODV, AOMDV, DSDV, AODVMOD and AOMDVMOD using 802.15.4 MAC protocol was realized by varying the network density and the amount of connections. The metrics to assess the performance are given as follows:

### a) End to End Delay

It can be defined as the average time between packets sent and received by the destiny nodes. Such metric can be calculated by using the Equation 1:

$$D = \frac{1}{N} \sum_{i=1}^{s} (r_i - s_i) \quad (1)$$

Where $N$ is the number of successfully received packets, $i$ is unique packet identifier, $r_i$ is time at which a packet with unique id $i$ is received, $s_i$ is time at which a packet with unique id $I$ is sent and $D$ is measured in ms. [20, 21].

### b) Packet Delivery Ratio (PDR)

Such metric can be defined as the percentage of the data packets delivered to the destinations to those generated by the sources. In Equation 2 is showed how such metric can be defined:

$$P = \frac{1}{c} \sum_{f=1}^{e} \frac{R_f}{N_f} \quad (2)$$

Where $P$ is the fraction of successfully delivered packets, $c$ is the total number of flow or connections, $f$ is the unique flow id serving as index, $R_f$ is the count of packet received from $f$ and $N_f$ is the count of packets transmitted to $f$ [20, 21].

### c) Averaged energy consumption

Such metric is important to be evaluated due to the fact of the WSN nodes have limited lifetime due to use batteries.

Considering $E$ be the initial energy of a node and residual energy $E_{res}$ of a node at time $t$, can be calculated by using the Equation 3 [22]:

$$E_{res} = E - E_{con}(t) \quad (3)$$

Where $E_{res}$ is the residual energy; $E_{con}$ is the consumed energy. Total energy consumption of all nodes is measured, as presented in Equation 4 as the summation of all node's residual energy plus the product of initial energy and number of nodes [22]:

$$T_{Econ} = N * \text{Initial Energy} - E_{res} \quad (4)$$

Where $T_{Econ}$ is the total consumed energy; $N$ is the total number of nodes used in the scenario.

## 5. RESULTS AND ANALYSIS

To discuss more about the results obtained in the simulations, they are showed in the Figures as follows.

In the Figure 6 is presented the results about Packet Delivery Ratio (PDR). AOMDV and AOMDVMOD protocols overcome other protocols due to fact to have multiple paths among source node and destiny node. When a path is lost, another path already established can be used to deliver packets. It is possible to see that the modification made in the HELLO messages interval of the protocol AOMDV, resulting in the protocol AOMDVMOD, impacted directly in the results in all scenarios evaluated, where in the scenario 1 is possible to see a performance about 25% higher of AOMDVMOD when compared with AOMDV due to the fact of the reduction of the HELLO messages, number of packet's collision and drop packets are reduced as well.

DSDV proactive protocol, needs to maintain your routing tables always updated and periodically, even if nodes do not have packets to send. Due to this reason, DSDV waits for a certain period to update its routing tables. In this interval of time, if some node wants to send packets, such packets are put in the queue and when the queue is full, packets will be dropped. That is why the packet delivery ratio of the DSDV protocol is too low as can be seen in the Figure 6. AODV and AODVMOD obtain better results than DSDV. The reason of this behaviour is that in case of link failure, DSDV waits for update packets while AODV broadcasts immediately a route request.

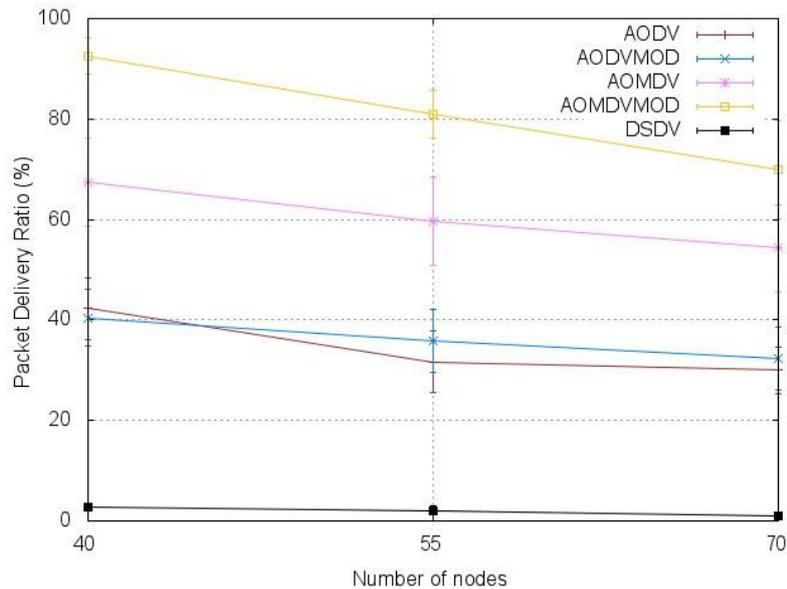

Figure 6. Results of Packet Delivery Ratio (%)

Figure 7 shows a comparison between the routing protocols on the basis of average end-to-end delay for different nodes number. Higher number of nodes is not necessary higher end-to-end delay values. Considering that there is no mobility in the nodes in our scenarios evaluated, the only problem that can increase delay will be battery. But in the simulations we saw that energy during the simulation time never ends. It is important to mention that DSDV have lower end to end delay, due to fact to have proactive behaviour and have fresh routes in its routing tables. AOMDV and AOMDVMOD have the best results with a little win of the AOMDVMOD and their results are near of the results obtained by DSDV. First, due to the fact to have multiple paths ("backup paths") to get destiny node in case of link failure and secondly, due to fact to send less HELLO messages broadcasts, reducing messages generated in the network avoiding collision and drop packets. AODV and AODVMOD routing protocols do not have satisfying results when compared with other protocols evaluated, because when there is a link failure, messages notifying such failure are necessary to inform the nodes in the network, generating more overhead in the network, increasing delay.

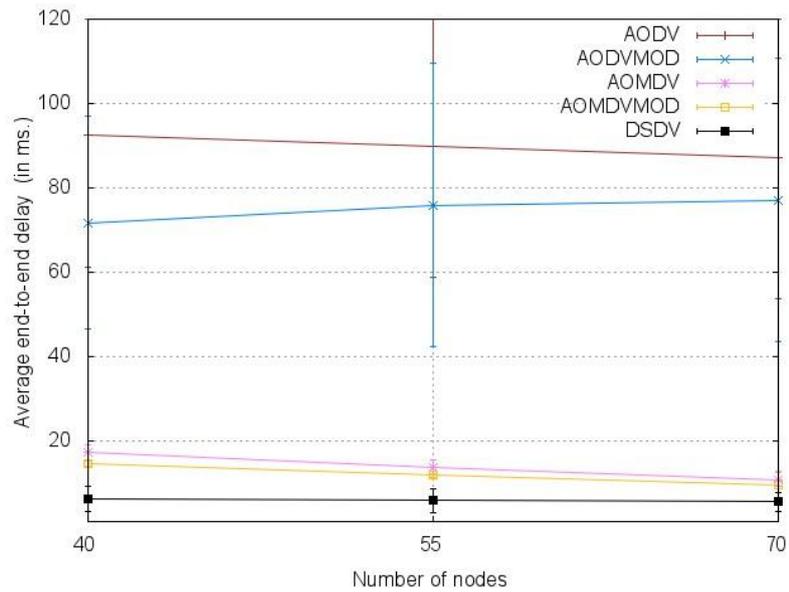

Figure 7. Results of End-to-end delay (in ms.)

Figure 8 shows the results comparing the average energy consumption of the routing protocols evaluated in this work. AOMDV and AOMDVMOD are the routing protocols that consume more energy. These protocols try to find multiple paths, using more resources and obviously consuming more energy. DSDV is the protocol that consumes less energy considering that it fills its routing tables without use techniques such as broadcast used by AODV and AODVMOD routing protocols, consuming more energy.

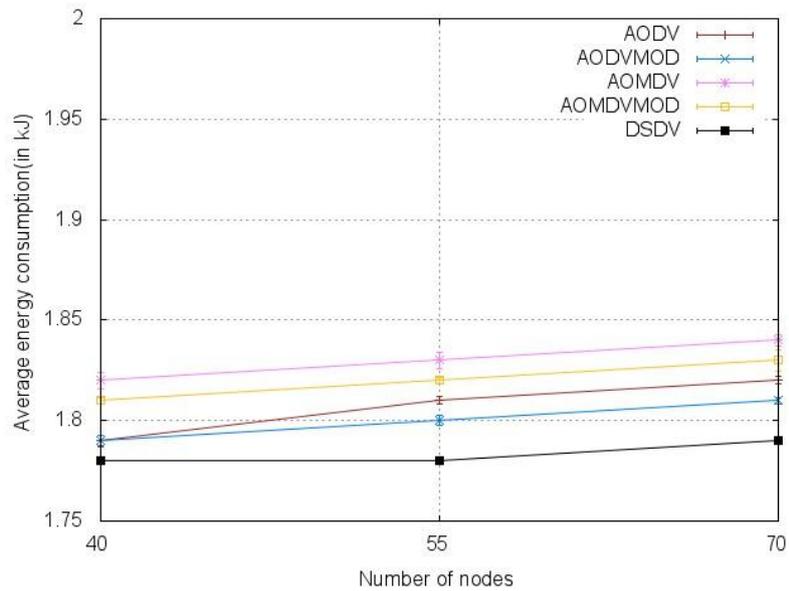

Figure 8. Results of Average Energy Consumption (in kJ.)

In the Table 4 we do a summary of the results, where the performance of the routing protocols are shown considering a score where 1 is worst result and 5 is the better result, considering the application aiming by us.

Table 4: Result's summary

| Metric/Protocol | AOMDV | AOMDVMOD | AODV | AODVMOD | DSDV |
|---|---|---|---|---|---|
| Packet Delivery Ratio | 4 | 5 | 3 | 3 | 1 |
| Average End-to-end Delay | 4 | 4 | 2 | 3 | 4 |
| Average Energy Consumption | 3 | 3 | 4 | 4 | 4 |
| Total point | 11 | 12 | 9 | 10 | 9 |

It is possible to see that considering Packet Delivery Ratio (PDR), AOMDVMOD is the best solution and have good results in other metrics evaluated. DSDV routing protocol has the worst result in PDR, but in the other metrics has good results. But considering the results obtained by DSDV protocol in PDR metric, and the application to be deployed, it shows that cannot be viable to be used. AODV and AODVMOD depending on the type of application, maybe they can be good routing protocols to be used in the network.

## 6. CONCLUSIONS

This work presented a study evaluating the feasibility of using Wireless Sensor Networks in a coffee crop through simulation using dimensions with a real area extracted from Google Earth and using characteristics of a coffee crop (such as spacing between coffee seeds). The results presented in this paper motivate the investigation of applicability of WSN in several areas. The results presented shows that AOMDV, overall, can be a good solution to use in a real deployment of a WSN.

As future works, the authors will deploy real sensors in the same area used in the simulations to evaluate the behavior of the routing protocols and to develop an Android software to manage the WSN deployed in a real environment to help farmers to have information in real-time.

## REFERENCES

[1] Akyildiz, Ian F. et al. 'Wireless sensor networks: a survey', Computer networks, v. 38, n. 4, p. 393–422, 2002.

[2] LOUREIRO, A. A. F.; NOGUEIRA, J. M. S.; RUIZ, L. B. 'Mini curso: Redes de sensores sem fio', Simpósio Brasileiro de Redes de Computadores (SBRC-2003).

[3] MITHILA, Nasrin Hakim. 'Performance analysis of DSDV, AODV and DSR in Wireless Sensor Network', International Journal of Advanced Research in Computer Science and Electronics Engineering (IJARCSEE), v. 2, n. 4, p. 395–404, 2013.

[4] OUNI, Sofiane and AYOUB, Z.T., "Precting Communication Delay and Energy Consumption for IEEE 802.15.4/Zigbee Wireless Sensor Networks. International Journal of Computer Networks & Communications (IJCNC) Vol.5, No.1, January 2013.

[5] PERKINS, C. E.; ROYER E. M.; DAS. S. R.; MARINA, M. K. "Performance comparison of two on-demand routing protocols for ad hoc networks", In: Proceedings on Personal Communications, 2001, pp. 16-28.

[6] PERKINS, C.; ROYER, E. M.; DAS, S. "Ad hoc On-Demand Distance Vector (AODV) Routing". Available in: http://www.faqs.org/rfcs/rfc3561.html. October 2008.

[7] MARINA, M. K.; DAS S. R., "Ad hoc On-demand Multipath Distance Vector Routing". ACM SIGMOBILE Mobile Computing and Communications Review, vol. 6-3, July 2002, pp. 92-93.

| [8] | Jacquet, P.; Muhlethaler, P.; Clausen, T.; Laouiti, A; Qayyum, A; Viennot, L., "Optimized link state routing protocol for ad hoc networks," *Multi Topic Conference, 2001. IEEE INMIC 2001. Technology for the 21st Century. Proceedings. IEEE International* , vol., no., pp.62,68, 2001. |
|---|---|
| [10] | N. Beijar, "Zone Routing Protocol (ZRP)", Ad Hoc Networking, Licentiate course on Telecommunications Technology, 2002. |
| [11] | D. Johnson, D. Maltz and Y-C. Hu, "The Dynamic Source Routing Protocol for Mobile Ad Hoc Networks," IETF Internet-Draft, Apr. 2003. |
| [12] | V. C. Patil, R. V. Biradar, R. R. Mudholkar, S. R. Sawant, "On-demand multipath routing protocols for mobile ad hoc networks issues and comparison", International Journal of Wireless Communication and Simulation, Vol. 2, No 1, pp. 21-38, 2010. |
| [13] | Indrani Das, D.K. Lobiyal and C.P. Kaiti. "Effect on Node Mobility on AOMDV Protocol in MANET". Interrnational Journal of Wireless & Mobile Networks (IJWMN), Vol. 6, No. 3, June 2014. |
| [14] | Chikha, H.B.; Makhlouf, A; Ghazel, W., "Performance analysis of AODV and DSR routing protocols for IEEE 802.15.4/ZigBee," *Communications, Computing and Control Applications (CCCA), 2011 International Conference on* , vol., no., pp.1,5, 3-5 March 2011. |
| [15] | Ramiro Liscano, John Khalil Jacoub, Anand Dersingh, Jinfu Zheng, Martin Helmer, Charles Elliott, and Ali Najafizadeh. 2011. Network performance of a wireless sensor network for temperature monitoring in vineyards. In *Proceedings of the 8th ACM Symposium on Performance evaluation of wireless ad hoc, sensor, and ubiquitous networks* (PE-WASUN '11). ACM, New York, NY, USA, 125-130. |
| [16] | SANTOS, I. M., DOTA, M. A., CUGNASCA, C. E., Visão Geral da Aplicabilidade de Redes de Sensores Sem Fio no Monitoramento Agrícola no estado de Mato Grosso. Anais do Congresso Brasileiro de Agricultura de Precisão 2010, Ribeirão Preto/SP, 2010. |
| [17] | Verona, A.B. 2010. Simulação e análise de redes de sensores sem fio aplicadas à viticultura. Dissertation-MSc in computer science. v548s. State University of Maringá - Brazil, Graduate Program in Computer Science. |
| [18] | Oliveira, F.D.M.; Semente, R.S.; Melo, T.A.C.; Salazar, A.O., "QoS analysis of routing protocols in wireless sensor networks in the monitoring of wind farms," *Instrumentation and Measurement Technology Conference (I2MTC) Proceedings, 2014 IEEE International* , vol., no., pp.1059,1064, 12-15 May 2014 |
| [19] | Information Sciences Institute (2014). NS-2 Network Simulator (version 2.34). Available at http://nsnam.isi.edu/nsnam/index.php/Main Page. |
| [20] | MITHILA, N. H. "Performance Analysis of DSDV, AODV and DSR in Wireless Sensor Network". International Journal of Advanced Research in Computer Science and Electronics Engineering (IJARCSEE), vol. 2, Issue 4, April 2013. |
| [21] | Nital Mistry, Devesh C Jinwala, Member, IAENG and Mukesh Zaveri. "Improving AODV Protocol against Blackhole Attacks", in Proc. IMECS'10, vol. 2, March 2010. |
| [22] | Shivashankar, Varaprasad G., Suresh H.N. Devaraju G. and Jayanthi G. Performance metrics Evaluation of Routing Protocols in MANET. International Journal of Advance Research in Computer and Communication Engineering. Vol.2, Issue 3, March 2013. |

**Authors**

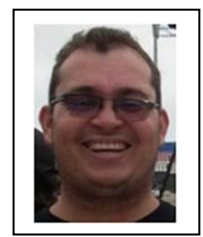

Ederval Pablo Ferreira da Cruz received the graduation in Data Processing from the University Center of Espirito Santo (UNESC), Brazil, in 2001 and the M.Sc degree from the Federal University of State of Rio de Janeiro (UNIRIO), Brazil, in 2012. Actually, he is a Ph.D special student at Federal University of Espirito Santo (UFES). He is currently professor of the Federal Institute of Espirito Santo (IFES) - Campus Itapina. His research interests include vehicular networks, wireless sensor networks, software-defined networks, Data Center Networks and Cloud Computing.

Luis Eduardo Gottardo is an undergraduate student at Federal Institute of Espirito Santo (IFES) - Campus Itapina. Currently, he has interests in precision agriculture using wireless sensor networks.

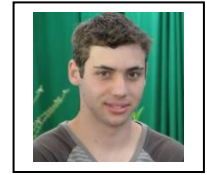

Franciele Rossini Pereira is an undergraduate student at the Federal Institute of Espirito Santo (IFES) – Campus Itapina. Currently she has interests in Research Development and Extension towards precision agriculture to hang on family farms.

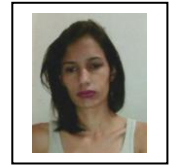

Vinicius de Souza Oliveira is an undergraduate student at Federal Institute of Espirito Santo (IFES) - Campus Itapina. Currently, he has interests in precision agriculture using wireless sensor networks.

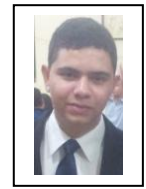

Lucas Cellim Pereira, in an undergraduate at Federal Institute of Espírito Santo (IFES) – Campus Itapina in agronomy college course. Currently, he has interests in mineral nutrition of plants, fertilization where is developing a software and in the precision agriculture.

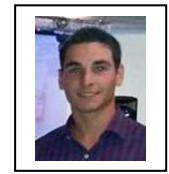